\documentclass[12pt]{article}
\usepackage{amsmath}
\usepackage{amsthm}
\usepackage{amscd}
\usepackage{epsfig}
\usepackage{fullpage}
\usepackage{natbib}
\usepackage{setspace}
\usepackage{amsfonts,bbm}

\newcommand{\bm}[1]{\mbox{\boldmath $#1$}}
\newcommand{\mb}[1]{\mathbf{#1}}

\newcommand{\Var}[0]{\mbox{Var}}
\newtheorem{prop}{Proposition}[section]
\newtheorem{remark}[prop]{Remark}

\begin{document}

\title{Importance Tempering}
\author{Robert Gramacy \& Richard Samworth\\ 
  Statistical Laboratory\\
  University of Cambridge\\
  \{bobby, rjs57\}@statslab.cam.ac.uk \and 
  Ruth King \\
  CREEM \\
  University of St Andrews \\
  ruth@mcs.st-and.ac.uk
}

\maketitle

\doublespacing

\begin{abstract}
  \noindent
  Simulated tempering (ST) is an established Markov chain Monte Carlo
  (MCMC) method for sampling from a multimodal density $\pi(\theta)$.
  Typically, ST involves introducing an auxiliary variable $k$ taking
  values in a finite subset of $[0,1]$ and indexing a set of tempered
  distributions, say $\pi_k(\theta) \propto \pi(\theta)^k$.  In this
  case, small values of $k$ encourage better mixing, but samples from
  $\pi$ are only obtained when the joint chain for $(\theta,k)$
  reaches $k=1$.  However, the entire chain can be used to estimate
  expectations under $\pi$ of functions of interest, provided that
  importance sampling (IS) weights are calculated.  Unfortunately this
  method, which we call importance tempering (IT), can disappoint.
  This is partly because the most immediately obvious implementation
  is na\"ive and can lead to high variance estimators.  We derive a
  new optimal method for combining multiple IS estimators and prove
  that the resulting estimator has a highly desirable property related
  to the notion of effective sample size.  We briefly report on the
  success of the optimal combination in two modelling scenarios
  requiring reversible-jump MCMC, where the na\"ive approach fails.

\end{abstract}

\section{Introduction}
\label{sec:intro}

Markov chain Monte Carlo (MCMC) algorithms, in particular
Metropolis--Hastings (MH) 
and Gibbs Sampling (GS),
are by now the most widely used methods for simulation--based
inference in Bayesian statistics.  The beauty of MCMC is its
simplicity.  Very little user input or expertise is required in order
to establish a Markov chain whose stationary distribution is
proportional to $\pi(\theta)$, for $\theta \in \Theta \subseteq
\mathbb{R}^d$.
As long as the chain is irreducible, the theory of Markov chains
guarantees that sample averages computed from this realisation will
converge in an appropriate sense to their expectations under $\pi$.
However, 
difficulties can arise when $\pi$ has isolated modes, between which
the Markov chain moves only rarely.  In such cases convergence is
slow, meaning that often infeasibly large sample sizes are needed to
obtain accurate estimates.

New MCMC algorithms have been proposed to improve mixing.  Two related
algorithms are Metropolis--coupled MCMC (MC$^3$)
\citep{geyer:1991,hukushima:1996} and simulated tempering (ST)
\citep{marinari:1992,geyer:1995}.  Both are closely related to the
optimisation technique of simulated annealing (SA) \citep{kirk:1983}.
SA works with a set of {\em tempered} distributions $\pi_k(\theta)$
indexed by an inverse--temperature parameter $k \in [0,\infty)$.  One
popular form of tempering is called ``powering up'', where
$\pi_k(\theta) \propto \pi(\theta)^k$.  Small values of $k$ have the
effect of flattening/widening the peaks and raising troughs in $\pi_k$
relative to $\pi$.  

In MC$^3$ and ST we define a {\em temperature ladder} $1 = k_1 > k_2 >
\ldots > k_m \geq 0$, and call the $k_i$ its {\em rungs}.  Both MC$^3$
and ST involve simulating from the set of $m$ tempered densities
$\pi_{k_1}$, \dots, $\pi_{k_m}$.  MC$^3$ runs $m$ parallel MCMC
chains, one at each temperature, and regularly proposes swaps of
states at adjacent rungs $k_i$ and $k_{i+1}$.  Usually, samples are
only saved from the ``cold distribution'' $\pi_{k_1}$.  In contrast,
ST works with a ``pseudo--prior'' $p(k_i)$ and uses a single chain to
sample from the joint distribution, which is proportional to
$\pi_k(\theta) p(k)$.  Again, it is only at iterations $t$ for which
$k^{(t)}=1$ that the corresponding realisation of $\theta^{(t)}$ is
retained.  ST has an advantage over MC$^3$ in that only one copy of
the process $\{\theta^{(t)}:t =1,\ldots,T\}$ is needed---rather than
$m$---so the chain uses less storage and also has better mixing
\citep{geyer:1991}.  The disadvantage is that it needs a good choice
of pseudo--prior.  For further comparison and review, see
\cite{jasra:steph:holm:2007} and \cite{iba:2001}.  

Both MC$^3$ and ST suffer from inefficiency because they discard all
samples from $\pi_k$ for $k \ne 1$.  The discarded samples could be
used to estimate expectations under $\pi$ if they were given
appropriate importance sampling (IS) weights.
For an inclusive review of IS and related methods see \citet[][Chapter
2]{liu:2001}.  Moreover, it may be the case that an IS estimator
constructed with samples from a tempered distribution has smaller
variance than one based on a sample of the same size from $\pi$.  As a
simple motivating example, let $\pi(\theta) = N(\theta|\mu,\sigma^2)$,
and consider estimating $\mu = \mathbb{E}_{\pi}(\theta)$ by IS from a
tempered distribution $\pi_k(\theta) \propto \pi(\theta)^k$.  A
straightforward calculation shows that the value of $k$ which
minimises the variance of the IS estimator is
\begin{equation}
k^* = \left\{\begin{array}{cl}
1/2 & \mbox{if } \mu= 0 \\
\frac{3}{2} + \Bigl(\frac{\sigma}{\mu}\Bigr)^2 - 
\frac{1}{2}\Bigl\{1 + 8\Bigl(\frac{\sigma}{\mu}\Bigr)^2 + 
4\Bigl(\frac{\sigma}{\mu}\Bigr)^4\Bigr\}^{1/2} & \mbox{otherwise.}
\end{array}\right. \label{eq:k}
\end{equation}
Note that $k^* \in (1/2,1)$ for all $\mu$ and $\sigma^2$.  Moreover,
one can compute (numerically) $k^- = k^-(\sigma/\mu) < k^*$ such that
for all $k \in (k^-,1)$, the variance of the IS estimator
$\hat{\mu}_k$ based on samples from $\pi_k$ is smaller than that of
one based on a sample of the same size from~$\pi$.  However,
$\mathrm{Var}(\hat{\mu}_k) \rightarrow \infty$ as $k \rightarrow 0$
for all $\mu$ and $\sigma^2$.  Table \ref{t:kminus} gives $k^*$ and
$k^-$ for various values of $\sigma/\mu$.
\begin{table}[ht!]
\centering
\vspace{0.2cm}
\begin{tabular}{c|ccccc}
$\sigma/\mu$ & 1/16 & 1/4 & 1 & 4 & 16 \\
\hline
$k^*$ & 1.00 & 0.95 & 0.70 & 0.52 & 0.50 \\
$k^-$ & 0.99 & 0.89 & 0.42 & 0.18 & 0.16 \\ 
\end{tabular}
\caption{Values of $k^*$ and $k^-$ for various values of $\sigma/\mu$.}
\label{t:kminus}
\end{table}

Therefore, there is a trade-off in the choice of tempered IS
proposals.  On the one hand, low inverse--temperatures $k$ in ST can
guard against missing modes of $\pi$ with large support by encouraging
better mixing {\em between} modes, but can yield very inefficient (IS)
estimators overall.  On the other hand, ``lukewarm'' temperatures $k$,
especially $k \in (1/2,1)$, can yield more efficient estimators {\em
  within} modes than those obtained from samples at $k=1$.

\cite{jennison:1993} was the first to suggest using a single tempered
distribution as a proposal in IS, and
\cite{neal:1996,neal:2001,neal:2005} has since written several papers
combining IS and tempering.  Indeed, in the discussion of the 1996
paper on {\em tempered transitions}, Neal writes ``simulated tempering
allows data associated with $p_i$ other than $p_0$ [the cold
distribution] to be used to calculate expectations with respect to
\dots \ $p_0$ (using an importance sampling estimator)''\footnote{A
  similar note is made in the 2001 paper with regard to {\em annealed
    importance sampling}.}.  It is this natural extension that we call
{\em importance tempering} (IT), with IMC$^3$ defined similarly.
Given the work of the above-mentioned authors, and the fact that
calculating importance weights is relatively trivial, it may be
surprising that successful IT and IMC$^3$ applications have yet to be
published.  \cite{liu:2001} comes close in proposing to augment ST
with dynamic weighting \citep{wong:1997} and in applying the
Wang--Landau algorithm \citep{atchade:2007} to ST.

This paper addresses why the straightforward methodology described
above has tended not to work well in practice, primarily due to a lack
of a principled way of combining the importance weights collected at
each temperature to obtain an overall estimator.  If we are interested
in estimating $\mathbb{E}_{\pi}\{h(\theta)\}$, one way to do this is
with
\begin{align}
\label{Eq:hath}
\hat{h} &= W^{-1} \sum_{t=1}^T w(\theta^{(t)},k^{(t)})h(\theta^{(t)}), 
& \mbox{where} &&
W = \sum_{t=1}^T w(\theta^{(t)},k^{(t)}),
\end{align}
and $w(\theta,k) = \pi(\theta)/\pi(\theta)^k = \pi(\theta)^{1-k}$.
Observe that this estimator is of the form $\hat{h} = \sum_{i=1}^m
\lambda_i \hat{h}_i$, where $0 \leq \lambda_i \leq \sum_{i=1}^m
\lambda_i = 1$, with $\lambda_i = W^{-1} \sum_{t=1}^T
w(\theta^{(t)},k^{(t)})\mathbb{I}_{\{k^{(t)} = k_i\}}$, and where
each $\hat{h}_i$ is an IS estimator of $\mathbb{E}_{\pi}\{h(\theta)\}$
constructed using only the observations at the inverse--temperature
$k_i$.  We show how to improve this estimator by choosing
$\lambda_1,\ldots,\lambda_m$ to maximise the \emph{effective sample
  size} (see next paragraph), which approximately corresponds to
minimising the variance of~$\hat{h}$
\citep[][Section~2.5.3]{liu:2001}.  For the applications that we have
in mind, it is important that our estimator can be constructed without
knowledge of the normalising constants of
$\pi_{k_1},\ldots,\pi_{k_m}$.
It is for this reason
that methods motivated by the \emph{balance heuristic} 
\citep{veach:1995,owen:2000,madras:1999}
cannot be applied.

The notion of \emph{effective sample size} plays an important role in
the study of IS estimators.  Suppose we are interested in estimating
$\mathbb{E}_{\pi}\{h(\theta)\}$ using a vector of observations
$\bm{\theta} = (\theta^{(1)},\ldots,\theta^{(T)})$ from a
density~$\pi'$.  Define the vector of importance weights $\mathbf{w}
\equiv \mathbf{w}(\bm{\theta}) = (w(\theta^{(1)}),\ldots,w(\theta^{(T)}))$,
where $w(\theta) = \pi(\theta)/\pi'(\theta)$.  Following
\citet[Section~2.5.3]{liu:2001} we define the \emph{effective sample
size} by
\begin{equation}
\mathrm{ESS}\bigl(\mathbf{w}(\bm{\theta})\bigr) \equiv
\mathrm{ESS}(\mb{w}) = \frac{T}{1 +
  \mathrm{cv^2}(\mathbf{w})}, \label{eq:essw}
\end{equation}
where $\mathrm{cv}^2(\mathbf{w})$ is the \emph{coefficient of variation}
of the weights, given by
\begin{align*}
\mathrm{cv^2}(\mathbf{w}) &= \frac{\sum_{t=1}^T(w(\theta^{(t)}) -
  \bar{w})^2}{(T-1) \bar{w}^2}, &\mbox{where} &&
\bar{w} &= T^{-1} \sum_{t=1}^T w(\theta^{(t)}).  
\end{align*}
This should not be confused with the concept of \emph{effective sample
  size due to autocorrelation} \citep{kass:1998} (due to serially
correlated samples from a Markov chain).  This latter notion is
discussed briefly in Section~\ref{sec:discuss}.


Observe that the swap operations in MC$^3$ require that the state
space $\Theta$ be common for all $m$ tempered distributions. This is
not a requirement for ST, as the state stays fixed when changes in
temperature are proposed.  Thus applying MC$^3$ is less
straightforward in (Bayesian) model selection/averaging problems which
typically involve trans--dimensional Markov chains as in
reversible--jump MCMC (RJMCMC) \citep{gree:1995}, though it is
possible \citep{jasra:bio:2007}.  Since RJMCMC algorithms are
particularly prone to slow mixing, and hence are an excellent source
of applications of our idea (as illustrated in
Section~\ref{sec:rjmcmc}), the rest of the paper will focus on IT.
Most of our results apply equally to IMC$^3$ by ignoring the
pseudo--prior.

The outline of the paper is as follows.  
In Section~\ref{sec:it} 
we derive the optimal convex combination of multiple IS estimators,
and show how this estimator has a particularly attractive property
with regard to its effective sample size.  In Section \ref{sec:rjmcmc}
we briefly report on the effectiveness of optimal IT, and the poor
performance of the na\"ive approach, on several real and synthetic
examples.  Section~\ref{sec:discuss} concludes with a discussion.

\section{Importance tempering}
\label{sec:it}



The {\em simulated tempering} (ST) \citep{geyer:1995} algorithm is an
application of MH on the product space of parameters and
inverse--temperatures.  That is, samples are obtained from
the joint chain $\pi(\theta,k) \propto \pi(\theta)^k p(k)$.
This is only possible if $\pi(\theta)^k$ is integrable, but
H\"older's inequality may be used to show that this is indeed the
case provided that $\mathbb{E}_{\pi}(\|\theta\|^{\frac{1-k}{k} +
  \delta}) < \infty$ for some $\delta > 0$, where $\|\cdot\|$
denotes the Euclidean norm.  
The success of ST depends crucially on the ability of the Markov chain
frequently to: (a) visit high temperatures (low $k$) where the
probability of escaping local modes is high; (b) visit $k=1$ to obtain
samples from $\pi$.  The algorithm can be tuned by: (i.)  adjusting
the number and location of the rungs of the temperature ladder; or
(ii.)  adjusting the pseudo-prior $p(k)$.
\citet{geyer:1995} give some automated ways of adjusting the spacing
of the rungs of the ladder. \cite{iba:2001} reviews similar techniques
from the physics literature. A recent alternative---and very
promising---approach involves the Wang--Landau algorithm
\citep{atchade:2007}.
However, many authors prefer to rely on defaults, e.g.,
\begin{equation}
 \;\;\;\;\;
k_i = \left\{ \begin{array}{cl}
(1+\Delta_k)^{1-i} & \mbox{geometric spacing}\\
\{1+\Delta_k (i-1)\}^{-1} & \mbox{harmonic spacing}
\end{array} \right. \;\;\;\;\ i=1,\dots,m.
\end{equation}
The rate parameter $\Delta_k>0$ can be problem specific.  
Motivation for such default spacings is outlined by \citet[][Chapter
10:~pp.~213 \& 233]{liu:2001}.  Geometric spacing, or uniform spacing
of $\log(k_i)$, is also advocated by \cite{neal:1996,neal:2001}. 

Once a suitable ladder has been chosen, the goal is typically to
choose the pseudo--prior so that the posterior over temperatures is
uniform.  The best way to accomplish this is to set $p(k_i) = 1/Z_i$,
where $Z_i = \int_\Theta \pi(\theta)^{k_i} d\theta$ is the normalising
constant in $\pi_{k_i} = \pi^{k_i}/Z_i$, which is generally unknown.
So while normalising constants are not a prerequisite for ST, it can
certainly be useful to know them.  We follow the suggestions of
\cite{geyer:1995} in setting the pseudo--prior by a method that
roughly approximates the $Z_i$ in two--stages: first by stochastic
approximation \citep{kush:1997}, and then by observation counts
accumulated through pilot runs. To some extent, a non-uniform
posterior on the temperatures is less troublesome in the context of IT
than ST.  So long as the chain still visits the heated temperatures
often enough to get good mixing in $\Theta$, and if the ESS of the IS
estimators at some temperature(s) is not too low, useful samples can
be obtained without ever visiting the cold distribution.


\subsection{A new optimal way to combine IS estimators}
\label{sec:lambdas}

ST provides us with $\{(\theta^{(t)},k^{(t)}): t = 1,\ldots,T\}$,
where $\theta^{(t)}$ is an sample from $\pi_{k^{(t)}}$.  Write
$\mathcal{T}_i = \{t: k^{(t)} = k_i\}$ for the index set of
observations at the $i^{\mbox{\tiny th}}$ temperature, and let $T_i =
|\mathcal{T}_i|$.  Let the vector of observations at the
$i^{\mbox{\tiny th}}$ temperature collect in $\bm{\theta}_i =
(\theta_{i1},\dots,\theta_{iT_i})$, so that
$\{\theta_{ij}\}_{j=1}^{T_i}\sim \pi_{k_i}$.  Similarly, the vector of
IS weights at the $i^{\mbox{\tiny th}}$ temperature is $\mathbf{w}_i =
\mathbf{w}_i(\bm{\theta}_i) =
(w_i(\theta_{i1}),\ldots,w_i(\theta_{iT_i}))$, where $w_i(\theta) =
\pi(\theta)/\pi_{k_i}(\theta)$.

Each vector $\bm{\theta}_i$ can be used to construct an IS
estimator of $\mathbb{E}_{\pi}\{h(\theta)\}$ by setting
\[
\hat{h}_i = \frac{\sum_{j=1}^{T_i} w_i(\theta_{ij}) h(\theta_{ij})}
{\sum_{j=1}^{T_i} w_i(\theta_{ij})} 
\equiv \frac{\sum_{j=1}^{T_i} w_{ij}h(\theta_{ij})}{W_i}.
\]
It is natural to consider an overall estimator of
$\mathbb{E}_{\pi}\{h(\theta)\}$ defined by a convex combination:
\begin{align}
\label{eq:hhatlambda}
\hat{h}_{\lambda} &= \sum_{i=1}^m \lambda_i \hat{h}_i,&
\mbox{where} && 0 \leq \lambda_i \leq \sum_{i=1}^m \lambda_i = 1.
\end{align}
Unfortunately, if $\lambda_1,\dots,\lambda_m$ are not chosen carefully,
$\Var(\hat{h}_\lambda)$, can be nearly as large as the largest
$\Var(\hat{h}_i)$ \citep{owen:2000}.  Notice that ST is recovered as a
special case when $\lambda_1=1$ and $\lambda_2 = \cdots = \lambda_m = 0$.
It may be tempting to choose $\lambda_i = W_i/W$, where $W =
\sum_{i=1}^m W_i$, recovering the estimator in Eq.~(\ref{Eq:hath}).
This can lead to a very poor estimator, even compared to ST, which is
demonstrated empirically in Section~\ref{sec:rjmcmc}.

Observe that we can write
\begin{equation}
\hat{h}_{\lambda} = \sum_{i=1}^m \sum_{j=1}^{T_i}
w_{ij}^{\lambda}h(\theta_{ij}), \label{eq:wlambda}
\end{equation}
where $w_{ij}^{\lambda} = \lambda_iw_{ij}/W_i$.  Let
$\mathbf{w}^{\lambda} =
(w_{11}^\lambda,\ldots,w_{1T_1}^\lambda,w_{21}^\lambda,\ldots,w_{2T_2}^\lambda,
\ldots,w_{m1}^\lambda,\ldots,w_{mT_m}^\lambda)$.  Attempting to choose
$\lambda_1,\dots,\lambda_m$ to minimise $\Var(\hat{h}_\lambda)$
directly can be difficult.  In the balance heuristic,
\citet{veach:1995} explore combinations of IS estimators of the form
(\ref{eq:wlambda}), where $w_i(\theta) = \pi(\theta)/g_i(\theta)$ for
a family of proposal densities $g_i$, with
\begin{equation}
  \lambda_{ij} = \frac{c_i g_i(\theta_{ij})}{\sum_{r=1}^m c_r g_r(\theta_{ij})},
\label{eq:bh}
\end{equation}
and where $0 \leq c_i \leq \sum_{i=1}^m c_i = 1$ is the proportion of
samples taken from $g_i$. 
It turns out
that this is equivalent to IS with the mixture proposal
$\tilde{\pi}(\theta) = \sum_{r=1}^m c_r g_r(\theta)$:
\begin{align}
\hat{h}_{\mbox{\tiny bal}} 
&\equiv \frac{1}{T}\sum_{t=1}^T w(\theta_t) h(\theta_t), &
\mbox{where} &&
w(\theta) &= \frac{\pi(\theta)}{\sum_{r=1}^m c_r g_r(\theta)}.
\label{eq:bhe}
\end{align}
The balance heuristic has since been generalised by \cite{owen:2000};
it was reinvented by \cite[][Section 4]{madras:1999} in the context of
applied probability.

Note that due to the denominator in the definition of $w(\theta)$ in
Eq.~(\ref{eq:bhe}), the $g_i$ must be normalised densities.  This
precludes us from using the balance heuristic with $g_i \propto
\pi_{k_i}$.  When MCMC is necessary to sample from $\pi$, the
normalisation constant of $\pi$, and therefore $\pi_{k_i}$, is
generally unknown.  
The method also requires evaluations of $\pi_{k_i}(\theta^{(t)})$,
$i=1,\dots,m$, at all $T$ rounds, an $O(mT)$ operation that
trivialises any computational advantage ST has over MC$^3$.  Instead,
we consider maximising the ESS of $\hat{h}_\lambda$ in
(\ref{eq:hhatlambda}).
\begin{prop}
\label{thm:lambdastar}
Among estimators of the form~(\ref{eq:hhatlambda}),
$\mathrm{ESS}(\mathbf{w}^{\lambda})$ is maximised by $\lambda =
\lambda^*$, where, for $i=1,\ldots,m$,
\begin{align*}
\lambda_i^* &= \frac{\ell_i}{\sum_{i=1}^m \ell_i}, & \mbox{and} && \ell_i &=
  \frac{W_i^2}{\sum_{j=1}^{T_i} w_{ij}^2}.
\end{align*}
\end{prop}
\begin{proof}
  Since $\sum_{i=1}^m \sum_{j=1}^{T_i} w_{ij}^\lambda = 1$, the
  problem of maximising the effective sample size is the same as
\begin{align*}
  \min_{\lambda_1,\dots,\lambda_m} \;\; & \sum_{i=1}^m
  \sum_{j=1}^{T_i} \left(\lambda_i \frac{w_{ij}}{W_i} -
    \frac{1}{T}\right)^2, && \mbox{subject to} & 0 & \leq \lambda_i
  \leq \sum_{i=1}^m \lambda_i = 1.
\end{align*}
The result then follows by a straightforward Lagrange multiplier argument.
\end{proof}
In the following discussion and in Remark~\ref{thm:ESS} below, we
assume that for $i=1,\ldots,m$, $T_i \geq 2$.  The efficiency of each
IS estimator $\hat{h}_i$ can be measured through
$\mathrm{ESS}(\mathbf{w}_i)$.  Intuitively, we hope that with a good
choice of $\lambda$, the ESS of $\hat{h}_{\lambda}$, given by
\[
\mathrm{ESS}(\mathbf{w}^{\lambda}) = \frac{T(T-1)}{T^2 \sum_{i=1}^m
  \lambda_i^2/\ell_i - 1},
\]
would be close to the sum over $i$ of the effective sample sizes of
$\hat{h}_i$, namely
\begin{equation}
\label{Eq:ESSwi}
\mathrm{ESS}(\mathbf{w}_i) = \frac{T_i(T_i-1)\ell_i}{T_i^2 - \ell_i}.
\end{equation}
The remark below shows that this is indeed the case for
$\hat{h}_{\lambda^*}$.
\begin{remark}
\label{thm:ESS}
We have
\[
\mathrm{ESS}(\mathbf{w}^{\lambda^*}) \geq \sum_{i=1}^m \mathrm{ESS}(\mathbf{w}_i) 
- \frac{1}{4} - \frac{1}{T}.
\]
\end{remark}
\begin{proof}
  Since $\mathrm{ESS}(\mathbf{w}_i) \leq T_i$, it follows
  from~(\ref{Eq:ESSwi}) that $\ell_i \leq T_i$.  Thus
\begin{align*}
  \mathrm{ESS}(\mathbf{w}^{\lambda^*}) =
  \frac{(1-T^{-1})\sum_{i=1}^m\ell_i} {1- \sum_{i=1}^m
  \frac{\ell_i}{T_i^2}} & \geq \Bigl(1-\frac{1}{T}\Bigr)
\biggl(1 + \frac{1}{T^2}\sum_{i=1}^m \ell_i\biggr)\sum_{i=1}^m\ell_i \\
&= \sum_{i=1}^m \ell_i - \frac{\sum_{i=1}^m \ell_i}{T}
\left(1 - \frac{\sum_{i=1}^m \ell_i}{T} \right)
- \frac{(\sum_{i=1}^m \ell_i)^2}{T^3} \\ &\geq \sum_{i=1}^m \ell_i -
\frac{1}{4} - \frac{1}{T},
\end{align*}
since $x(1-x)$ attains its maximum of $1/4$ at $x=1/2$ and $\sum
\ell_i \leq \sum T_i = T$.
\end{proof}

In practice we have found that this bound is slightly conservative and
that often it is the case that $\mathrm{ESS}(\mathbf{w}^{\lambda^*})
\geq \sum_{i=1}^m \mathrm{ESS}(\mathbf{w}_i)$.  Thus our
optimally--combined IS estimator has a highly desirable and intuitive
property in terms of its effective sample size.

\section{Empirical Results}
\label{sec:rjmcmc}



Here we briefly report on the success of optimal IT, relative to the
na\"ive approach and ST, on one simple example and two involving
RJMCMC.

\subsection{A simple mixture of normals}

Consider the following toy density 
$\pi$, a mixture of two normals:  
\begin{equation}
  \pi(\theta) = 0.6 N(\theta|\mu_1=-8, \sigma^2_1=0.5^2) +
  0.4 N(\theta|\mu_2=8, \sigma^2_2=0.9^2). \label{eq:normix}
\end{equation}
Table~\ref{Table:MSEComp2} summarises Kolmogorov--Smirnov distances
obtained under three IT estimators: ST ($\lambda_1 = 1$), na\"ive IT
($\lambda_i = W_i/W$) and the optimally--combined IT estimator
($\hat{h}_{\lambda^*}$).
\begin{table}[ht!]
\centering
\begin{tabular}{l|ccc} 
&& \multicolumn{2}{c}{K--S distance} \\
Method & 
$\mathrm{ESS}(\mb{w}^\lambda)$ & mean & var \\ 
\hline
ST & 2535 & 0.0938 & $8.5\times10^{-4}$ \\ 
na\"ive IT & 17779 &  0.0849 &  $1.4\times10^{-4}$ \\
$\hat{h}_{\lambda^*}$ & 22913 & 0.0836 & $5.2\times10^{-5}$ \\
\hline
$\sum_i \mathrm{ESS}(\mb{w}_i)$ & 22910 
\end{tabular}
\caption{Summary of K--S distances to the true mixture of normals (\ref{eq:normix}) 
  for ST ($\lambda_1 = 1$), na\"ive IT ($\lambda_i = W_i/W$), the
  optimally--combined IT estimator ($\hat{h}_{\lambda^*}$). We used
  100 repeated samples of size $10^5$, with tempered RWM proposals. }
\label{Table:MSEComp2}
\end{table}
Observe that the optimally--combined IT estimator has both the largest
ESS and the smallest variance of the three estimators, and that
$\mathrm{ESS}(\mb{w}^{\lambda^*}) > \sum_i \mathrm{ESS}(\mb{w}_i)$.
Na\"ive IT improves upon ST in this example, but has higher variance
than $\hat{h}_{\lambda^*}$.  

\subsection{Bayesian treed Gaussian process models}

Bayesian treed models extend classification and regression tree
(CART) models \citep{brei:1984}, by putting a prior on the tree
structure. 
We focus on the implementation of \cite{gra:lee:2008} who fit Gaussian
Process (GP) models at the leaves of the tree, specify the tree prior
through a process that limits its depth, and then define the tree
operations {\em grow}, {\em prune}, {\em change}, and {\em swap}, to
allow inference to proceed by RJMCMC. The RJMCMC chain usually
identifies the correct {\em maximum a posteriori} (MAP) tree, but
consistently and significantly over estimates the posterior
probability of deep trees.

To guard against the transdimensional chain getting stuck in local
modes of the posterior,
\cite{gra:lee:2008} resorted regularly restarting the chain from the
null tree.
ST provides an alternative by increasing the rate of accepted tree
operations in higher temperatures.  In particular, we find that ST can
increase the rate of accepted {\em prune} operations by an order of
magnitude, thus enabling the chain to escape the local modes of deep
trees.  To demonstrate IT we fit a treed GP model with ST using a
geometric ladder with $m=40$ and $k_m = 0.1$ to two datasets first
explored by \cite{gra:lee:2008}: the 1-d motorcycle accident data and
2-d exponential data.  We refer to that paper for details about the
data and models.



For the motorcycle accident data the ST chain was run for $T=1.5\times
10^5$ iterations, where a total of $T_1=3732$ ($\approx T/m=3750$)
samples were obtained from the cold distribution.
That $\mathrm{ESS}(\mathbf{w}^{\lambda^*})=9338 \approx 2.5 T_1$ shows
the considerable improvement of IT over ST.  Moreover, we have
$\mathrm{ESS}(\mathbf{w}^{\lambda^*}) > \sum_i \mathrm{ESS}(\mb{w}_i) =
9334$. The na\"{i}ve combination $\lambda_i = \frac{W_i}{W}$ in
(\ref{Eq:hath}) yields $\mathrm{ESS}(\mathbf{w}^{\lambda}) = 285 <
\frac{1}{10} T_1$, undermining the very motivation of IT.
For the exponential data the ST chain was run for a total of
$T=5\times 10^5$ iterations.
A total of $T_1=12436$ ($\approx T/m=12500$) samples were obtained
from the cold distribution.  
We found that $\mathrm{ESS}(\mathbf{w}^{\lambda^*})=21778\approx 1.75
T_1$, illustrating how IT improves on ST. Moreover, we have
$\mathrm{ESS}(\mathbf{w}^{\lambda^*}) > \sum_i \mathrm{ESS}(\mb{w}_i)
= 21776$. The na\"{i}ve combination $\lambda_i = \frac{W_i}{W}$ in
(\ref{Eq:hath}) yields $\mathrm{ESS}(\mathbf{w}^{\lambda^*}) = 654
\approx \frac{1}{18} T_1$---worse than ST.

\subsection{Mark-Recapture-Recovery Data}

We now consider a Bayesian model selection problem with data relating
to the mark-recapture and recovery of shags on the Isle of May 
\citep{king:brooks:2002}.
The three demographic parameters of interest are: survival rates,
recapture rates and recovery rates.  The models considered for each of
the demographic parameters allowed a possible age-- and/or
time--dependence, where the time dependence was conditional on the age
structure of the parameters.  Typically, movement between the
different possible models---by adding/removing time dependence for a
given age group, or updating the age structure of the parameters---is
slow, with small acceptance probabilities. For further details of the
data, model structure, and RJMCMC algorithm see
\cite{king:brooks:2002}.

Using the same ST setup as above, we ran 
%
$T=10^7$ iterations and discarded the first 10\% as burn-in.  As with
the treed examples, higher temperatures yielded higher acceptance rates
and an order of magnitude better exploration of model space compared
to (untempered) RJMCMC.
A total of $T_1 = 248158 \ (\approx T/m = 225000)$ realisations were
obtained from the cold distribution.  By comparison, for optimal IT we
have $\mathrm{ESS}(\mathbf{w}^{\lambda^*}) = 612026 \approx 2.5 T_1$
and $\mathrm{ESS}(\mathbf{w}^{\lambda^*}) > \sum_i
\mathrm{ESS}(\mb{w}_i) = 612020$.  The corresponding na\"ive IT
approach (using $\lambda_i = \frac{W_i}{W}$) performed exceptionally
poorly, with $\mathrm{ESS}(\bf w^{\lambda})$ of only 5.43, due to a
few large weights obtained at hot temperatures.

\section{Discussion}
\label{sec:discuss}

This paper has addressed the inefficiencies and wastefulness of
simulated tempering (ST), and related algorithms that are designed to
improve mixing in the Markov chain using tempered distributions.  We
argued that importance sampling (IS) from tempered distributions can
produce estimators that are more efficient than ones based on
independent sampling, provided that the temperature is chosen
carefully.  This motivated augmenting the ST algorithm by calculating
importance weights to salvage discarded samples---a technique which we
have called {\em importance tempering} (IT).  This idea has been
suggested before, but to our knowledge little
exploration has been carried out for real, complex, applications.
%
We have derived optimal combination weights for the resulting
collection of IS estimators, which can be calculated even when the
normalisation constants of the tempered distributions are unknown.
The weights are essentially proportional to the effective sample size
(ESS) of the individual estimators, and we found that the resulting
combined ESS in this case would be approximately equal to their 
sum.

We note that the overall success of the optimal IT estimator depends
crucially on a successful implementation of ST, i.e., having a good
temperature ladder and pseudo--prior.  However, it is also important
to recognise that the optimal combination, as a resource--efficient
post-processing step, is equally applicable in other contexts, i.e.,
within MC$^3$, or even outside of the domain of tempered MCMC to
combine any collection IS estimators.  Sequential Monte Carlo samplers
\citep{delmoral:2006} may facilitate a natural
extension.  
%
We have illustrated IT on several examples which benefit from the
improved mixing ST provides.  For example, the optimal IT methodology
can increase the resulting ESS compared to retaining samples only from
the cold distribution by roughly a factor of
two.  

Since IT involves sampling from a Markov chain, ideally one would take
into account the serial correlation in the objective criteria for
combining the individual estimators.  The {\em effective sample size
  due to autocorrelation} is defined \citep{kass:1998} by
\begin{equation}
\mathrm{ESS}_\rho(\bm{\theta}) = \frac{T}{1 + 2\sum_{\ell=1}^{T-1}
  \hat{\rho}(\ell,\bm{\theta})}, \label{eq:essc}
\end{equation}
where $\hat{\rho}(\ell, \bm{\theta})$ is the sample autocorrelation in
$\bm{\theta}$ at lag $\ell$; thus for scalar $\theta$ we have that
$\hat{\rho}(\ell,\bm{\theta}) =
\hat{\gamma}(\ell,\bm{\theta})/\hat{\gamma}(0,\bm{\theta})$, where
$\hat{\gamma}(\ell,\bm{\theta}) = (T-\ell)^{-1}\sum_{t=1}^{T-\ell}
(\theta^{(t)} - \bar{\theta})(\theta^{(t+\ell)} - \bar{\theta})$, and
$\bar{\theta} = T^{-1}\sum_{t=1}^T
\theta^{(t)}$.  
The results from the previous section suggest that, when the
temperature ladder is fixed, a sensible heuristic might be to consider
combining the individual estimators with weights $\lambda_i^*$
proportional to product of $T_i^{-1}\mathrm{ESS}_\rho(\bm{\theta}_i)$
and $\mathrm{ESS}(\mathbf{w}_i)$, say.  However, when considering
modifications to the number ($m$) and spacing of inverse temperatures
$\mb{k} = \{k_1,\dots, k_m\}$, there is clearly a conflict of interest
between the two measures of effective sample size. Adding more inverse
temperatures near one may increase $\mathrm{ESS}(\mb{w}^{\lambda^*})$,
but may also increase autocorrelation in the marginal chain for $k$.
Therefore it may be sensible to factor $\mathrm{ESS}_\rho(\mb{k})$
into the objective as well.  Searching for temperature ladders that
maximise a hybrid of $\mathrm{ESS}$ and $\mathrm{ESS}_\rho$ would
represent a natural extension of this work.


\renewcommand{\baselinestretch}{1.5}\small\normalsize

\bibliography{mhis}
\bibliographystyle{jasa}

\end{document}